# Eigenforms and Quantum Physics
## by Louis H. Kauffman

**Dedication:** To Heinz on his 100-th birthday.

*But thought's the slave of life, and life time's fool;*
*And time, that takes survey of all the world,*
*Must have a stop.*

(Hotspur, Act 5, Scene 4, Henry the IVth, Part I, Shakespeare.)

**Abstract:** This essay is a discussion of the concept of eigenform and its relationship with the foundations of physics.

## I. Introduction

Our essay begins with explication of the notion of eigenform as pioneered by Heinz von Foerster in his papers [4, 5,6,7] and explored in papers of the author [11, 12, 22, 23]. In [5] The familiar objects of our existence can be seen as tokens for the behaviors of the organism, creating apparently stable forms.

Such an attitude toward objects makes it impossible to discriminate between the object as an element of a world and the object as a token or symbol that is simultaneously a process.

The notion of an eigenform is inextricably linked with second order cybernetics. One starts on the road to such a concept as soon as one begins to consider a pattern of patterns, the form of form or the cybernetics of cybernetics. Such concepts appear to loop around upon themselves, and at the same time they lead outward to new points of view. Such circularities suggest a possibility of transcending the boundaries of a system from within. When a circular concept is called into being, the boundaries turn inside out.

An object , in itself , is a symbolic entity, participating in a network of interactions, taking on its apparent solidity and stability from these interactions. We ourselves are such objects, we as human beings are "signs for ourselves", a concept originally due to the American philosopher C. S. Peirce [10]. Eigenforms are mathematical companions to Peirce's work.

In an observing system, what is observed is not distinct from the system itself, nor can one make a separation between the observer and the observed. The observer and the observed stand together in a coalescence of perception. From the stance of the observing system all objects are non-local, depending upon the presence of the system as a whole. It is within that paradigm that these models begin to live, act and enter into conversation with us.

The central metaphor of this paper is the temporal nexus where time is implicit, and time is explicit and keeping time. In the nexus there is neither form nor sign, motion nor time. Time, the measurement of time and time's indication all emerge at once from the nexus in the form of action that is embodied in **it**. The metaphor suggests that it is no accident that deeper physical reality is revealed when mere numerical time **t** is replaced by the time of the nexus **it**. The time of the nexus is at once flowing, beyond motion, an eigenform, a geometric operator and a discrete dynamics counting below where counting cannot go. The author hopes that Heinz will understand his attempt to say what cannot be said. The sound of the nexus is silence when it is truly read.

## II. Objects as Tokens for Eigenbehaviours

In his paper "Objects as Tokens for Eigenbehaviours" [5] von Foerster suggests that we think seriously about the mathematical structure behind the constructivist doctrine that *perceived worlds are worlds created by the observer*. At first glance such a statement appears to be nothing more than solipsism. At second glance, the statement appears to be a tautology, for who else can create the rich subjectivity of the immediate impression of the senses? In that paper he suggests that the familiar objects of our experience are the fixed points of operators. These operators *are* the structure of our perception. To the extent that the operators are shared, there is no solipsism in this point of view. It is the beginning of a mathematics of second order cybernetics.

Consider the relationship between an observer O and an "object" A. "The object remains in constant form with respect to the observer". This constancy of form does not preclude motion or change of shape. Form is more malleable than the geometry of Euclid. In fact, ultimately the form of an "object" is the form of the distinction that "it" makes in the space of our perception. In any attempt to speak absolutely about the nature of form we take the form of distinction for the form. (paraphrasing Spencer-Brown [3]). It is the form of distinction that remains constant and produces an apparent object for the observer. How can you write an equation for this? We write

$$O(A) = A.$$

The object **A** is a fixed point for the observer **O**. The object is an eigenform. We must emphasize that this is a most schematic description of the condition of the observer in relation to an object **A.** We record only that the observer as an actor (operator) manages to leave the (form of) the object unchanged. This can be a recognition of symmetry, but it also can be a description of how the observer, searching for an object, makes that object up (like a good fairy tale) from the very ingredients that are the observer herself.

And what about this matter of the object as a token for eigenbehaviour? This is the crucial step. We forget about the object and focus on the observer. We attempt to "solve" the equation **O(A) = A** with **A** as the unknown. Not only do we admit that the "inner" structure of the object is unknown, we adhere to whatever knowledge we have.

We can start anew from the dictum that the perceiver and the perceived arise together in the condition of observation. This is mutuality. Neither perceiver nor the perceived have priority over the other. A distinction has emerged and with it
a world with an observer and an observed. The distinction is itself an  eigenform.

### III. The Eigenform Model
We have seen how the concept of an object has evolved. The notion of a fixed object has become the notion of a process that produces the apparent stability of an object. This process can be simplified in modeling to become a recursive process where a rule or rules are applied time and time again. The resulting object is the fixed point or *eigenform* of the process, and the process itself is the *eigenbehaviour*.

In this way we have a model for thinking about object as token for eigenbehaviour. This model examines the result of a simple recursive process carried to its limit.
For example, suppose that

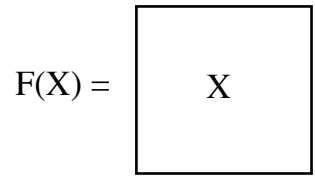

**Figure 1**

Each step in the process encloses the results of the previous step within a box. Here is an illustration of the first few steps of the process applied to an empty box X:

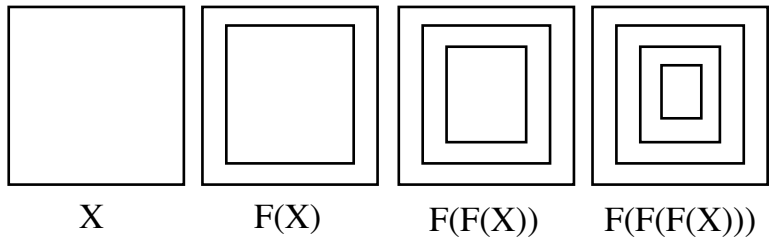

**Figure 2**

If we continue this process, then successive nests of boxes resemble one another, and in the limit of infinitely many boxes, we find that the infinite nest of boxes is invariant under the addition of one more surrounding box. Hence this infinite nest of boxes is a fixed point for the recursion. In other words, if X denotes the infinite nest of boxes, then

$$X = F(X).$$

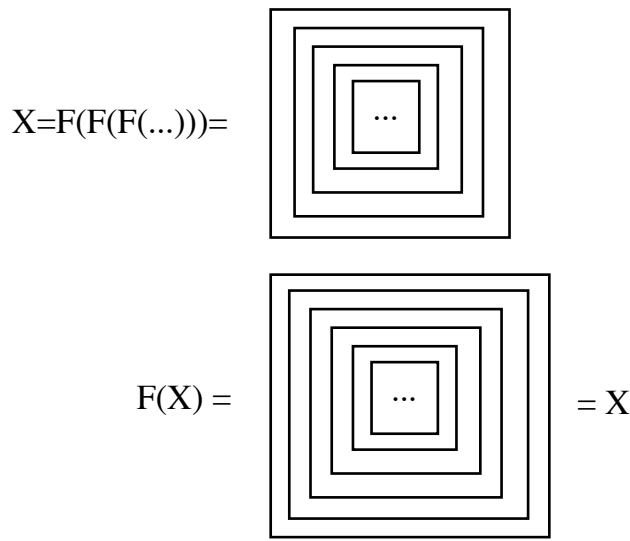

**Figure 3**

This equation is a description of a state of affairs. The form of an infinite nest of boxes is invariant under the operation of adding one more surrounding box.

In the process of observation, we interact with ourselves and with the world to produce stabilities that become the objects of our perception. These objects, like the infinite nest of boxes, often go beyond the specific properties of the world in which we operate. We make an imaginative leap to complete such objects to become tokens for eigenbehaviours. It is impossible to make an infinite nest of boxes. We do not make it. We *imagine* it. And in imagining that infinite nest of boxes, we arrive at the eigenform.

Sometimes one stylizes the structure by indicating where the eigenform **X** reenters its own indicational space by an arrow or other graphical device. See the picture below for the case of the nested boxes.

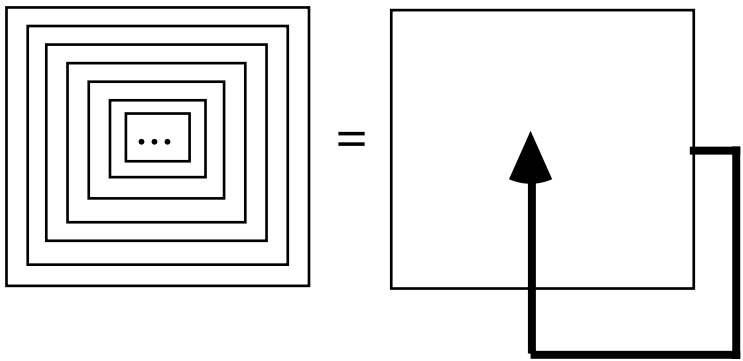

**Figure 4**

An object is an amphibian between the symbolic and imaginary world of the mind and the complex world of personal experience. The object, when viewed as process, is a dialogue between these worlds. The object when seen as a sign for itself, or in and of itself, is imaginary. The perceiving mind is itself an eigenform of its own perception.

**IV. The Square Root of Minus One**

The purpose of this section is to place **i**, square root of minus one, and its algebra in the context of eigenform and reflexivity. We describe a process point of view for complex numbers. Think of the oscillatory process generated by

$$R(x) = -1/x.$$

The fixed point would satisfy
$$i = -1/i$$
and multiplying, we get that
$$ii = -1.$$
On the other hand the iteration of **R** yields

$$+1, \ R(1) = -1, \ R(R(1)) = +1, -1, +1, ....$$

The square root of minus one is a perfect example of an eigenform that occurs in a new and wider domain than the original context in which its recursive process arose. The process has no fixed point in the original domain.

Looking at the oscillation between **+1** and **-1**, we see that there are naturally two viewpoints that we denote by **[+1,-1]** and **[-1,+1]**. These viewpoints correspond to whether one regards the oscillation at time zero as starting with **+1** or with **-1**.

We shall let **I{+1,-1}** stand for an undisclosed alternation or ambiguity between **+1** and **-1** and call **I{+1,-1}** an *iterant*. There are two i*terant views*: **[+1,-1]** and **[-1,+1]**.

Given an iterant **[a,b]**, we can think of **[b,a]** as the same process with a shift of one time step.

These two iterant views, seen as points of view of an alternating process, will become the square roots of negative unity, **i** and **-i.**

We introduce a *temporal shift operator* $\eta$ such that

$$[a,b]\eta = \eta [b,a] \text{ and } \eta \eta = 1$$

for any iterant **[a,b]**, so that concatenated observations can include a time step of one-half period of the process

$$...abababab... \ .$$

We combine iterant views term-by-term as in

$$[a,b][c,d] = [ac,bd].$$

We now define **i** by the equation

$$\mathbf{i} = [1,-1]\eta \ .$$

This makes **i** both a value and an operator that takes into account a step in time.

We calculate

$$\mathbf{ii} = [1,-1]\eta \ [1,-1]\eta \ = [1,-1][-1,1]\eta \ \eta \ = [-1,-1] = -1.$$

Thus we have constructed the square root of minus one by using an iterant viewpoint. **i** represents a discrete oscillating temporal process and it is an eigenform for **R(x) = -1/x,** participating in the algebraic structure of the complex numbers.

**The Temporal Nexus**
*We take as a matter of principle that the usual real variable **t** for time is better represented as **it** so that time is seen to be a process, an observation and a magnitude all at once.*
This principle of "imaginary time" is justified by the eigenform approach to the structure of time and the structure of the square root of minus one.

An example of the use of the Temporal Nexus, consider the expression $x^2 + y^2 + z^2 + t^2$, the square of the Euclidean distance of a point **(x,y,z,t)** from the origin in Euclidean four-dimensional space. Now replace **t** by i**t**, and find

$$x^2 + y^2 + z^2 + (it)^2$$
$$= x^2 + y^2 + z^2 - t^2,$$

the squared distance in hyperbolic metric for special relativity. By replacing **t** by its process operator value **it** we make the transition to the physical mathematics of special relativity.

**V. Quantum Physics, Eigenvector, Eigenvalue and Eigenform**

The reader should recall the Temporal Nexus from Section IV. *We take as a matter of principle that the usual real variable **t** for time is better represented as **it** so that time is seen to be a process, an observation and a magnitude all at once.*

The reader should recall that a *vector* is a quantity with magnitude and direction, often pictured as an arrow in the plane or in three dimensional space.

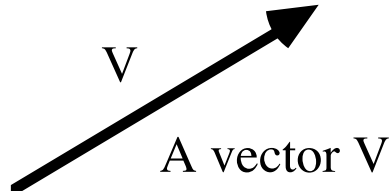

A vector V

**Figure 5**

In quantum physics [11], the state of a physical system is modeled by a vector in a high-dimensional space, called a Hilbert space. As time goes on the vector rotates in this high dimensional space. Observable quantities correspond to (linear) operators **H** on these vectors **v** that have the property that the application of **H** to **v** results in a new vector that is a multiple of **v** by a real factor **λ.** (An operator is said to be linear if **H(av +w) = aH(v) + H(w)** for vectors v and w, and any number a. Linearity is usually a simplifying assumption in mathematical models, but it is an essential feature of quantum mechanics.)

In symbols this has the form

$$\mathbf{Hv} = \lambda \mathbf{v}.$$

One says that **v** is an *eigenvector* for the operator **H**, and that **λ** is the *eigenvalue*.

The theory of eigenforms is a sweeping generalization of quantum mechanics shifting eigenvectors to eigenforms.

This is a reversal of epistemology, a complete turning of the world upside down. Eigenform has tricked us into considering the world of our experience and we find that it is our world, generated by our actions. The world has become objective through the self-generated stabilities of those actions.

## VI. The Wave Function in Quantum Mechanics

One can regard a wave function such as $\psi(x,t) = \exp(i(kx - wt))$ as containing a micro-oscillatory system with the special synchronizations of the iterant view $i = [+1,-1]\eta$. It is these synchronizations that make the big eigenform of the exponential $\psi(x,t)$ work correctly with respect to differentiation, allowing it to create the appearance of rotational behaviour, wave behaviour and the semblance of the continuum.

One can blend the classical geometrical view of the complex numbers with the iterant view by thinking of a point that orbits the origin of the complex plane, intersecting the real axis periodically and producing, in the real axis, a periodic oscillation in relation to its orbital movement in the higher dimensional space.

## VII. Time Series and Discrete Physics

We have just reformulated the complex numbers and expanded the context of matrix algebra to an interpretation of **i** as an oscillatory process and matrix elements as combined spatial and temporal oscillatory processes (in the sense that **[a,b]** is not affected in its order by a time step, while **[a,b]η** includes the time dynamic in its interactive capability, and 2 x 2 matrix algebra is the algebra of iterant views **[a,b] + [c,d]η** ).

We now consider elementary discrete physics in one dimension. Consider a time series of positions
$x(t), t = 0, \Delta t, 2\Delta t, 3\Delta t, \ldots$ . We can define the velocity $v(t)$ by the formula
$$v(t) = (x(t+\Delta) - x(t))/\Delta t = Dx(t)$$
where **D** denotes this discrete derivative. In order to obtain $v(t)$ we need at least one tick $\Delta t$ of the discrete clock. Just as in the iterant algebra, we need a time-shift operator to handle the fact that once we have observed $v(t)$, the time has moved up by one tick.

Thus we shall add an operator **J** that in this context accomplishes the time shift:

$$x(t)J = Jx(t+\Delta t).$$

We then *redefine* the derivative to include this shift:

$$Dx(t) = J(x(t+\Delta) - x(t))/\Delta t .$$

The result of this definition is that a successive observation of the form **x(Dx)** is distinct from an observation of the form **(Dx)x**. In the first case, we observe the velocity and then **x** is measured at **t + Δt**. In the second case, we measure **x** at **t** and then measure the velocity.

We measure the difference between these two results by taking a commutator **[A,B] = AB - BA** and we get the following formula where we write **Δx = x(t+ Δt) - x(t).**

$$[x,(Dx)] = x(Dx) - (Dx)x$$
$$= (J/\Delta t)(x(t+\Delta t) - x(t))^2$$
$$= J (\Delta x)^2/\Delta t$$

This final result is worth marking:

$$[x,(Dx)] = J (\Delta x)^2/\Delta t.$$

From this result we see that the commutator of **x** and **Dx** will be constant if $(\Delta x)^2/\Delta t = K$ is a constant. For a given time-step, this means that $(\Delta x)^2 = K \Delta t$ so that $\Delta x = +\sqrt{K \Delta t}$ or $-\sqrt{K \Delta t}$. This is a Brownian process with diffusion constant equal to **K.**

## VIII. Epilogue and Simplicity
Finally, we arrive at the simplest place.
Time and the square root of minus one are inseparable in the temporal nexus. The square root of minus one is a symbol and algebraic operator for the simplest oscillatory process.
As a symbolic form, **i** is an eigenform satisfying the equation
$$i = -1/i.$$
One does not have an increment of time all alone as in classical **Δt.** One has **iΔt**, a combination of an interval and the elemental dynamic that is time. With this understanding, we can return to the commutator for a discrete process and use **iΔt** for the temporal increment.

We found that discrete observation led to the commutator equation

$$[x, Dx] = J\,(\Delta x)^2/\Delta t$$

which we will simplify to

$$[q,\ p/m] = (\Delta x)^2/\Delta t.$$

taking **q** for the position **x** and **p/m** for velocity, the time derivative of position.

Understanding that **Δt** should be replaced by **iΔt**, and that, by comparison with the physics of a process at the Planck scale one can take

$$(\Delta x)^2/\Delta t = \hbar/m,$$

we have

$$[q,\ p/m] = (\Delta x)^2/i\,\Delta t = -i\,\hbar/m,$$

whence

$$[p,q] = i\hbar,$$

and we have arrived at Heisenberg's fundamental relatiionship between position and momentum. This mode of arrival is predicated on the recognition that only **i Δt** represents a true interval of time. In the notion of time there is an inherent clock or an inherent shift of phase that is making a synchrony in our ability to observe, a precise dynamic beneath the apparent dynamic of the observed process. Once this substitution is made, *once the correct imaginary value is placed in the temporal circuit, the patterns of quantum mechanics appear*.

The problem that we have examined in this paper is the problem to understand the nature of quantum mechanics. In fact, we hope that the problem is seen to disappear the more we enter into the present viewpoint. A viewpoint is only on the periphery. The iterant from which the viewpoint emerges is in a superposition of indistinguishables, and can only be approached by varying the viewpoint until one is released from the particularities that a point of view contains.

**References**
1. H. P. Barendregt, "The Lambda Calculus - Its Syntax and Semantics," North Holland Pub. (1981,1985).

2. H. Bortoft, The Whole - Counterfeit and Authentic, Systematics , vol. 9, No. 2, Sept. (1971), 43-73.

3. G. Spencer-Brown, "Laws of Form," George Allen and Unwin Ltd. (1969).

4. Heinz von Foerster, "Observing Systems," The Systems Inquiry Series, Intersystems Publications (1981).

5. Heinz von Foerster, Objects: tokens for (eigen-) behaviors, in "Observing Systems," The Systems Inquiry Series, Intersystems Publications (1981), pp. 274 - 285.

6. Heinz von Foerster, Notes on an epistemology for living things, in "Observing Systems," The Systems Inquiry Series, Intersystems Publications (1981), pp. 258 - 271.

7. Heinz von Foerster, On constructing a reality, in "Observing Systems," The Systems Inquiry Series, Intersystems Publications (1981), pp. 288 - 309.

8. L. H. Kauffman, Self-reference and recursive forms, Journal of Social and Biological Structures (1987), 53-72.

9. L. H. Kauffman, Knot logic, in "Knots and Applications", ed. by L. H. Kauffman, World Scientific Pub. Co. (1995), pp. 1-110.

10. L. H. Kauffman, The mathematics of Charles Sanders Peirce, in Cybernetics and Human Knowing, Volume 8, No. 1-2, (2001), pp. 79-110.

11. Louis H. Kauffman, Eigenform, Kybernetes - The Intl J. of Systems and Cybernetics, Vol. 34, No. 1/2 (2005), Emerald Group Publishing Ltd, p. 129-150.

12. Louis H. Kauffman, Eigenforms - Objects as Tokens for Eigenbehaviors, Cybernetics and Human Knowing, Vol. 10, No. 3-4, 2003, pp. 73-90.

13. F.W. Lawvere, Introduction to "Toposes, Algebraic Geometry and Logic" Springer Lecture Notes on Mathematics Vol. 274 (1970), pp. 1-12.

14. B. B. Mandelbrot, "The Fractal Geometry of Nature", W. H. Freeman and Company (1977, 1982).

15. W. S. McCulloch, What is a number that a man may know it, and a man, that he may know a number?, in "Embodiments of Mind", MIT Press (1965), pp. 1-18.

16. B. Piechocinska, Physics from Wholeness, PhD. Thesis, Uppsala Universitet (2005).

17. J. J. Sakurai, Modern Quantum Mechanics, Benjamin/Cummings Publishing Company, Inc. (1985).

18. D. Scott, Relating theories of the lambda calculus, in "To H. B. Curry: Essays on Combinatory Logic, Lambda Calculus and Formalism", (P. Seldin and R. Hindley eds.), Academic Press (1980), pp. 403-450.

19. L. Wittgenstein, "Tractatus Logicus Philosophicus", Routledge and Kegan Paul Ltd, London and New York (1922).

20. Louis H. Kauffman, Non-Commutative Worlds, New Journal of Physics, Vol. 6, (2004), 173 (47 pages).

21. Patrick DeHornoy, "Braids and Self-Distributivity", Birkhaurser (2000).

22. Louis H. Kauffman, Reflexivity and Eigenform - The Shape of Process, Constructivist Foundations, Vol. 4, No. 3, July 2009, pp. 121-137.

23. Louis H. Kauffman, Reflexivity and Foundations of Physics, In "Search for Fundamental Theory - The VIIth Intenational Symposium Honoring French Mathematical Physicist Jean-Pierre Vigier, Imperial

College, London, UK, 12-14 July 2010", AIP - American Institute of Physics Pub., Melville, N.Y., pp.48-89.